\documentclass[twocolumn,showpacs,showkeys,preprintnumbers,prd,superscriptaddress,nofootinbib]{revtex4-1}

\usepackage{amsmath}
\usepackage{amsfonts}
\usepackage{amssymb}
\usepackage{graphicx}
\usepackage{color}
\usepackage[]{hyperref}
\usepackage{cleveref}
\usepackage{mathtools}

\Crefname{equation}{Eq.}{Eqs.}
\Crefname{figure}{Fig.}{Figs.}
\Crefname{section}{Sec.}{Secs.}
\Crefrangeformat{equation}{Eqs.~(#3#1#4)--(#5#2#6)}

\begin{document}
\title{Reconstructing the distortion function of non-local cosmology: a model-independent approach}

\author{Salvatore Capozziello}
\email{capozziello@na.infn.it}
\affiliation{Dipartimento di Fisica ``E. Pancini", Universit\`a di Napoli ``Federico II", Via Cinthia 21, 80126 Napoli, Italy}
\affiliation{Scuola Superiore Meridionale, Largo S. Marcellino 10, 80138 Napoli, Italy}
\affiliation{Istituto Nazionale di Fisica Nucleare (INFN), Sezione di Napoli, Via Cinthia 21, 80126 Napoli, Italy}

\author{Rocco D'Agostino}
\email{rocco.dagostino@unina.it}
\affiliation{Scuola Superiore Meridionale, Largo S. Marcellino 10, 80138 Napoli, Italy}
\affiliation{Istituto Nazionale di Fisica Nucleare (INFN), Sezione di Napoli, Via Cinthia 21, 80126 Napoli, Italy}

\begin{abstract}

We consider  non-local modifications of General Relativity  given by a distortion function in terms of the inverse of the d'Alembert operator. The inclusion  of these terms is motivated by the possibility of reproducing the current accelerated expansion  of the Universe starting from non-local gravity. 
In particular, we propose a model-independent method, based on current observations, to reconstruct  the shape of the distortion function, without resorting to a specific cosmological history. We describe a numerical procedure based on  Pad\'e polynomials that allows us to study the time evolution of the auxiliary scalar fields introduced to localize non-local gravity action. Thus, adopting suitable boundary conditions, we reconstruct the form of the non-local Lagrangian and infer the best analytical approximation of the numerical outcome. 
Furthermore, the distortion function turns on during the matter-dominated era and its effects are delayed to the most recent cosmic times. This provides a natural explanation for the late cosmic acceleration, avoiding any fine-tuning problem of the cosmological constant. 
Finally, we compare our predictions with previous findings based on enforcing the standard $\Lambda$CDM background in the reconstruction process.

\end{abstract}

\maketitle

\section{Introduction}

 General Relativity (GR) has been confirmed over time by several  experimental and observational tests resulting in a self-consistent and robust description of gravitational phenomena \cite{Will:2014kxa,LIGOScientific:2016aoc}. 
Nevertheless, several theoretical limitations and shortcomings, affecting GR, emerged during the years at the high and low-energy scales \cite{Koyama:2015vza,Berti:2015itd,Ishak:2018his,DAgostino:2020dhv,Vagnozzi:2023nrq,DAgostino:2023cgx}. In the ultra-violet regime, the main difficulty is the lack of consistency with Quantum Mechanics.
In the infrared regime, the standard cosmological scenario based on GR, i.e., the $\Lambda$CDM model, is characterized by the presence of unknown sources  dominating the matter-energy budget of the Universe \cite{Planck18,Perivolaropoulos:2021jda,DAgostino:2023tgm}. In such a picture, the vacuum energy, under the form of  cosmological constant $\Lambda$, dominates the late comic epoch, giving rise to the accelerated expansion of the today observed Universe  \cite{Riess98,Perlmutter99,Peebles:2002gy}. 
The energy density of $\Lambda$, inferred from cosmological estimates, however, differs greatly from the quantum predictions, leading to the so-called \emph{fine-tuning issue} \cite{Weinberg89,Padmanabhan:2002ji,DAgostino:2022fcx}.

The dark energy problem indicates the possibility that GR may be not enough to explain all  gravitational phenomenology, so several alternative theories have been proposed in the last years with the aim to extend or modify Einstein's gravity \cite{Clifton:2011jh,Capozziello:2011et,Cai:2015emx,Nojiri:2017ncd,Capozziello:2019cav,Capozziello:2022uak}. A common strategy is to consider additional (global or local) degrees of freedom that might be able to overcome the shortcomings of GR by means of new fields \cite{Linder:2010py,Nojiri:2010wj,DAgostino:2018ngy,BeltranJimenez:2017tkd,DAgostino:2019wko,DAgostino:2019hvh,Anagnostopoulos:2021ydo,DAgostino:2021vvv,Bajardi:2022tzn,DAgostino:2022tdk}. 
Along this line, an interesting approach is to encode modifications of the gravitational sector through the presence of non-local terms, being non-locality one of the main features of Quantum Mechanics \cite{Deser:2007jk,Calcagni:2007ru,Nojiri:2007uq,Koivisto:2008xfa,Maggiore:2014sia,Dirian:2014ara,Nersisyan:2016hjh,Capozziello:2022rac}. This would be a step towards a full theory of Quantum Gravity. Specifically, higher-order terms, involved in theories  extending GR to  high-energy regimes, show that renormalizable Lagrangians can be actually obtained in four dimensions \cite{Calcagni:2010ab,Modesto:2011kw,Briscese:2012ys}.  In this case, instabilities resulting from higher-order derivative operators may be avoided by relaxing the locality principle \cite{Biswas:2011ar,Buoninfante:2018xiw,Calcagni:2018lyd}. It is worth noticing that non-local corrections to the gravitational Lagrangian can emerge as conserved quantities generated by Noether symmetries \cite{Capozziello:2021krv, Acunzo, Libro}.  In general,  non-local terms typically appear in loop corrections of the effective actions of quantum gravity \cite{Barvinsky:1985an,Goroff:1985th,Shapiro:2008sf}. Furthermore, some classes of non-local theories of gravity have been used to address black hole and Big Bang singularities \cite{Li:2015bqa,Modesto:2017sdr}. 

Among all the non-local modified gravity proposals, an intriguing possibility is to consider a gravitational action containing the term $R f(\Box^{-1}R)$, where $R$ is the Ricci curvature and $\Box^{-1}$ is the inverse of the d'Alembert operator. In its original formulation \cite{Deser:2007jk}, this model was shown to properly account for the transitions between different cosmological epochs.
However, solar system tests revealed that some experimental constraints are violated due to the lack of a screening mechanism for non-local effects at short distances \cite{Belgacem:2017cqo}. Later on, the model was improved to overcome these issues and provides a viable scenario capable of reproducing the expansion history of the standard $\Lambda$CDM paradigm \cite{Deser:2019lmm} as well as the dynamics of clusters of galaxies \cite{Filippo1}. Furthermore, it can be useful to fix  cosmological tensions in the framework of $\Lambda$CDM \cite{Filippo2}. More recently, this improved framework was also taken into account to examine the viability of theoretical models within the context of bouncing cosmology \cite{Jackson:2021mgw} and of possible astrophysical tests at galactic scales \cite{Vesna1,Vesna2}.

Motivated by the aforementioned considerations, in the present study, we consider the non-local Lagrangian analyzed in \cite{Deser:2019lmm}. 
Although, in principle, to obtain the time behaviour of the non-local term, one needs to specify the pressure and the energy density of the cosmic fluid, however, an alternative approach could be adopted in order to reduce possible biases induced by imposing a specific background history given by a chosen cosmological model. 
In particular, in this work, we propose a model-independent procedure to infer the distortion function encoding the non-local modification of Einstein's gravity. We demonstrate that the accelerated expansion of the Universe could be obtained with no need to introduce dark energy or other exotic fluids, avoiding \emph{de facto} the fine-tuning problem proper of the cosmological constant. For this purpose, we make use of rational Pad\'e approximations to parametrize the cosmic expansion and express the Hubble function in terms of kinematic quantities that do not depend on any postulated cosmological scenario \cite{Wei:2013jya,Aviles:2014rma,Capozziello:2017nbu,Nesseris:2013bia,Capozziello:2020ctn}. Hence, a numerical procedure based on recent constraints from cosmic observations will be able to provide the shape of the distortion function and, thus, to pinpoint the Lagrangian of the theory.

The paper is structured as follows. In \Cref{sec:theory}, we analyze the cosmological features of the non-local gravitational scenario. Then, in \Cref{sec:reconstruction}, we present the methodology to reconstruct the shape of the non-local distortion function. In particular, we first describe the cosmographic technique based on Pad\'e polynomials and the Bayesian analysis for constraining the cosmographic parameters from current observations.  Then, using our numerical bounds, we investigate the dynamical behaviour of the auxiliary fields emerging from the localization of the action. Thus, we provide the analytical expression that best approximates the numerical solution for the distortion function.
In \Cref{sec:results}, we discuss  results and compare them with previous predictions in the literature. Finally, \Cref{sec:conclusions} is dedicated to the summary of our findings and concluding remarks. Throughout this paper, we adopt units such that $c=\hbar=1$.

\section{Non-local cosmology}
\label{sec:theory}

A non-local modification to the Einstein-Hilbert action is \cite{Deser:2019lmm}
\begin{equation}
S=\frac{1}{16\pi G} \int d^4x\, R\big[1+f(Y)\big]\sqrt{-g}\,,
\label{action_nonlocal}
\end{equation}
where $g$ is the determinant of the metric tensor $g_{\mu\nu}$, and
\begin{align}
X& = \Box^{-1}R\,, \label{eq:X}\\ 
Y&= \Box^{-1}\left(g^{\mu\nu}\partial _\mu X \partial_\nu X\right)  \label{eq:Y}\,,
\end{align}
with $\Box\equiv \nabla^\mu\nabla_\mu$ being the d'Alembert operator. Here, $f(Y)$ is known as the {\it distortion function}, encoding the effects of the   fields $X$ and $Y$.

It is possible to localize the action \eqref{action_nonlocal} by introducing two  auxiliary scalar fields, $U$ and $V$, treated as Lagrange multipliers:
\begin{equation}
S=\frac{1}{16\pi G} \int d^4x\, \Big\{R\left[1+U+f(Y)\right]+g^{\mu\nu}B_{\mu\nu}\Big\}\sqrt{-g}\,,
\label{action_localized}
\end{equation}
where, for convenience, we introduce the tensor
\begin{equation}
B_{\mu\nu}\equiv \partial_\mu X\partial_\nu U +\partial_\mu Y \partial_\nu V+V\partial_\mu X \partial_\nu X \,.
\end{equation}
Varying the action \eqref{action_localized} with respect to $X$ and $Y$ provides  the dynamical equations for the fields $U$ and $V$, respectively:
\begin{align}
U&=-2\Box^{-1}\nabla_\mu(V \nabla^\mu X)\,, \label{eq:U}\\
V&=\Box^{-1}\left(R\, \dfrac{df}{dY}\right) . \label{eq:V}
\end{align}
We interestingly note that $X$, $Y$, $U$ and $V$ are considered as independent scalar fields and the gravitational action is regarded as local. Within this approach, being the auxiliary fields treated as Lagrange multipliers, \Cref{eq:X,eq:Y,eq:U,eq:V} obtained from the variational principle correspond to the Klein-Gordon equations for each field. 

It is also worth stressing that all the auxiliary scalars obey retarded boundary conditions and are vanishing, together with their first derivatives with respect to time, when evaluated at the initial value surface\footnote{The initial value surface, in our case, refers to early cosmic times (see \Cref{sec:reconstruction}).}. This feature ensures the absence of extra degrees of freedom, preventing the introduction of ghosts \cite{Deser:2013uya}.
Moreover,  the auxiliary fields propagate along the characteristic curves of the d'Alembert scalar, thus the sound speed coincides with the speed of light, avoiding  issues typical of several modified theories of gravity \cite{Sawicki:2015zya}. 

From the variation of the action \eqref{action_localized} with respect to $g^{\mu\nu}$, we obtain the field equations
\begin{multline}
\left(G_{\mu\nu}+g_{\mu\nu}\Box-\nabla_\mu \nabla_\nu\right)\left[1+U+f(Y)\right]+B_{(\mu\nu)}  \\
-\frac{1}{2}g_{\mu\nu}g^{\alpha\beta}B_{\alpha\beta} =8\pi G T_{\mu\nu}\,,
\label{eq:FE}
\end{multline}
where $G_{\mu\nu}\equiv R_{\mu\nu}-\frac{1}{2}g_{\mu\nu}R$ is the Einstein tensor, and $T_{\mu\nu}$ is the matter energy-momentum tensor of the cosmic fluid. The round parentheses denote symmetrized indices, such that $B_{(\mu\nu)}=\frac{1}{2}(B_{\mu\nu}+B_{\nu\mu})$.

With the purpose of studying the cosmological features of the above non-local model, let us assume the spatially flat Friedman-Lema\^itre-Robertson-Walker (FLRW) metric:
\begin{equation}
ds^2=dt^2-a(t)^2\delta_{ij}dx^i dx^j\,,
\label{eq:metric}
\end{equation}
where $a(t)$ is the cosmic scale factor, normalized at the present time. Therefore, introducing the Hubble parameter $H(t)\equiv \dot{a}/a$, the d'Alembert operator reads 
\begin{equation}
\Box=\frac{d^2}{dt^2}+3H\frac{d}{dt}\,.
\end{equation}
In view of the above assumptions, from \Cref{eq:FE} one obtains the modified Friedman equations:
\begin{multline}
3H\left(H+\frac{d}{dt}\right)\left[1+U+f(Y)\right]+\frac{1}{2}\left(\dot{X}^2+\dot{X}\dot{U}+\dot{Y}\dot{V}\right)\\ 
=8\pi G\rho\,,
 \label{eq:Friedmann1}
\end{multline}\\
\vspace{-1.2cm}
\begin{multline}
-\left(3H^2+2\dot{H}+\frac{d^2}{dt^2}+2H\frac{d}{dt}\right)\left[1+U+f(Y)\right]  \\
+\frac{1}{2}\left(\dot{X}^2+\dot{X}\dot{U}+\dot{Y}\dot{V}\right)= 8\pi G p\,,
 \label{eq:Friedmann2}
\end{multline}
where $\rho$ and $p$ are, respectively, the density and the pressure of the fluid without the dark energy contribution. 
Moreover, combining \Cref{eq:Friedmann1,eq:Friedmann2}, we find
\begin{equation}
\left(6H^2+2\dot{H}+\frac{d^2}{dt^2}+5H\frac{d}{dt}\right)W(t)=8\pi G (\rho-p)\,,
\label{eq:master}
\end{equation}
where $W\equiv 1+U+f$. Solving \Cref{eq:master} for $W(t)$ leads to the distortion function and, thus, to the reconstruction of the non-local Lagrangian form. 

While this methodology was originally adopted  to reproduce the $\Lambda$CDM cosmology \cite{Deser:2019lmm}, in this paper, we shall present a novel approach  allowing us to find the time evolution of the auxiliary scalar fields, and then to infer the shape of the distortion function, with no \emph{a priori} assumptions on the underlying cosmology. We describe the reconstruction procedure  in the next section.

\section{The reconstruction method}
\label{sec:reconstruction}

The primary aim of the non-local model under consideration is to provide an alternative explanation to the current Universe acceleration  free from any fine-tuning problem of the cosmological constant. Therefore, we shall not enforce the $\Lambda$CDM expansion history, or any other, in order to minimize possible biases in the reconstruction of $f(Y)$. Rather, we seek to determine the distortion function in a model-independent way. Combining  analytical and numerical recipes,  our method relies on   what follows.
 
\subsection{Cosmography with Pad\'e polynomials}

A powerful method allowing for a model-independent reconstruction of modified gravity actions is offered by Pad\'e polynomials within the cosmographic framework \cite{Capozziello:2017ddd,Capozziello:2018aba,Capozziello:2022wgl,Capozziello:2022jbw}.
In particular, one can Taylor-expand the scale factor of the FLRW metric around the present time as 
\begin{equation}
a(t)=1+\sum_{k=1}^{\infty}\dfrac{1}{k!}\dfrac{d^k a}{dt^k}\bigg | _{t=t_0}(t-t_0)^k\,,
\label{eq:scale factor}
\end{equation}
where the coefficients of the expansion define the so-called cosmographic series \cite{Visser:2004bf,Cattoen:2007sk}:
\begin{equation}
H\equiv \dfrac{1}{a}\dfrac{da}{dt}\,,\quad q\equiv -\dfrac{1}{aH^2}\dfrac{d^2a}{dt^2}\,, \quad j \equiv \dfrac{1}{aH^3}\frac{d^3a}{dt^3}\,.
\end{equation}
Here, $q$ and $j$ are the deceleration and jerk parameters, respectively.
The present-day values of the cosmographic parameters can be used to find a kinematic expansion of the luminosity distance in terms of the redshift variable, $z\equiv a^{-1}-1$:
\begin{equation}
d_L(z)=\dfrac{z}{H_0}\bigg(1+\sum_{k=1}^\infty c_k z^k\bigg)\,,
\label{eq:luminosity distance}
\end{equation}
where the first three coefficients read \cite{Capozziello:2019cav}
\begin{subequations}
\begin{align}
&c_1=\dfrac{1}{2}(1-q_0)\,,		\\
&c_2=-\dfrac{1}{6}(1-q_0-3q_0^2+j_0)\,, \\
&c_3=\dfrac{1}{24}(2-2q_0-15q_0^2-15q_0^3+5j_0+10q_0j_0+s_0)\,.
\end{align}
\end{subequations}
Then, the Hubble parameter is obtained through the standard relation
\begin{equation}
H(z)=\left[\dfrac{d}{dz}\left(\dfrac{d_L(z)}{1+z}\right)\right]^{-1}\,.
\label{eq:Hubble rate}
\end{equation}

The standard cosmography  described above, albeit straightforward to implement, is affected by some problems due to the short convergence radius of the Taylor series, which  reflects the limited prediction power when handling data at $z>1$. On the other hand, the lack of very accurate high-redshift measurements leads to the struggle to constrain the high-order coefficients of the cosmographic expansion. 

A remarkable way to heal such issues is offered by Pad\'e polynomials, which can be used to construct rational approximations of cosmographic observables that are characterized by stable behaviours at large cosmological distances and extended convergence radii \cite{Aviles:2014rma}. Specifically, given the Taylor series of an arbitrary function of the redshift, $f(z)=\sum_{k=0}^\infty c_kz^k$, the Pad\'e approximation of the order $(n,m)$ of $f(z)$ is given by the ratio 
\begin{equation}
P_{n, m}(z)=\dfrac{\displaystyle{\sum_{i=0}^{n}a_{i} z^{i}}}{\displaystyle{\sum_{j=0}^{m}b_j z^{j}}}\,,
\end{equation}
where the unknown coefficients $a_i$ and $b_i$ are determined through the following system:
\begin{equation}
\left\{
\begin{aligned}
&a_i=\sum_{k=0}^i b_{i-k}\ c_{k} \ ,  \\
&\sum_{j=1}^m b_j\ c_{n+k+j}=-b_0\ c_{n+k}\ , \hspace{0.5cm} k=1,\hdots, m \ .
\end{aligned}
\right .
\end{equation}

In a previous study \cite{Capozziello:2020ctn}, we showed in detail the advantages of Pad\'e approximations in terms of convergence, stability and accuracy when used to frame the Universe's evolution. In view of discriminating among the infinite polynomial orders, we demonstrated that the (2,1) Pad\'e parametrization is the most suitable approximation able to provide accurate cosmographic results, being less prone to numerical error propagation due to the reduced number of free parameters. In particular, the (2,1) Pad\'e parametrization of the luminosity distance is given as
\begin{equation}
d_{2,1}(z)=\dfrac{1}{H_0}\bigg[\dfrac{z (6 (q_0-1) + (q_0 (8 + 3 q_0)-5 - 2 j_0 ) z)}{2 q_0 (3 + z + 3 q_0 z)-2 (3 + z + j_0 z) }\bigg]\,.
\label{eq:dL_Pade21}
\end{equation}

The above expression can be compared directly to observations in order to constrain the values of the cosmographic coefficients and describe the cosmic expansion without resorting to specific background models. 
As we are dealing with a model-independent description of the Universe's evolution, we choose here to neglect measurements that are acquired through the use of a fiducial cosmology. Thus, we adopt cosmic chronometers (CC) and type Ia Supernovae (SN) measurements, which represent robust and reliable datasets. In fact, they can be effectively adopted to reconstruct the Hubble expansion rate avoiding possible biases induced by the choice of the cosmological model. We refer to \Cref{sec:appendix} for the details on these datasets and the relative likelihood functions. 

\begin{figure}
    \centering
    \includegraphics[width=3.3in]{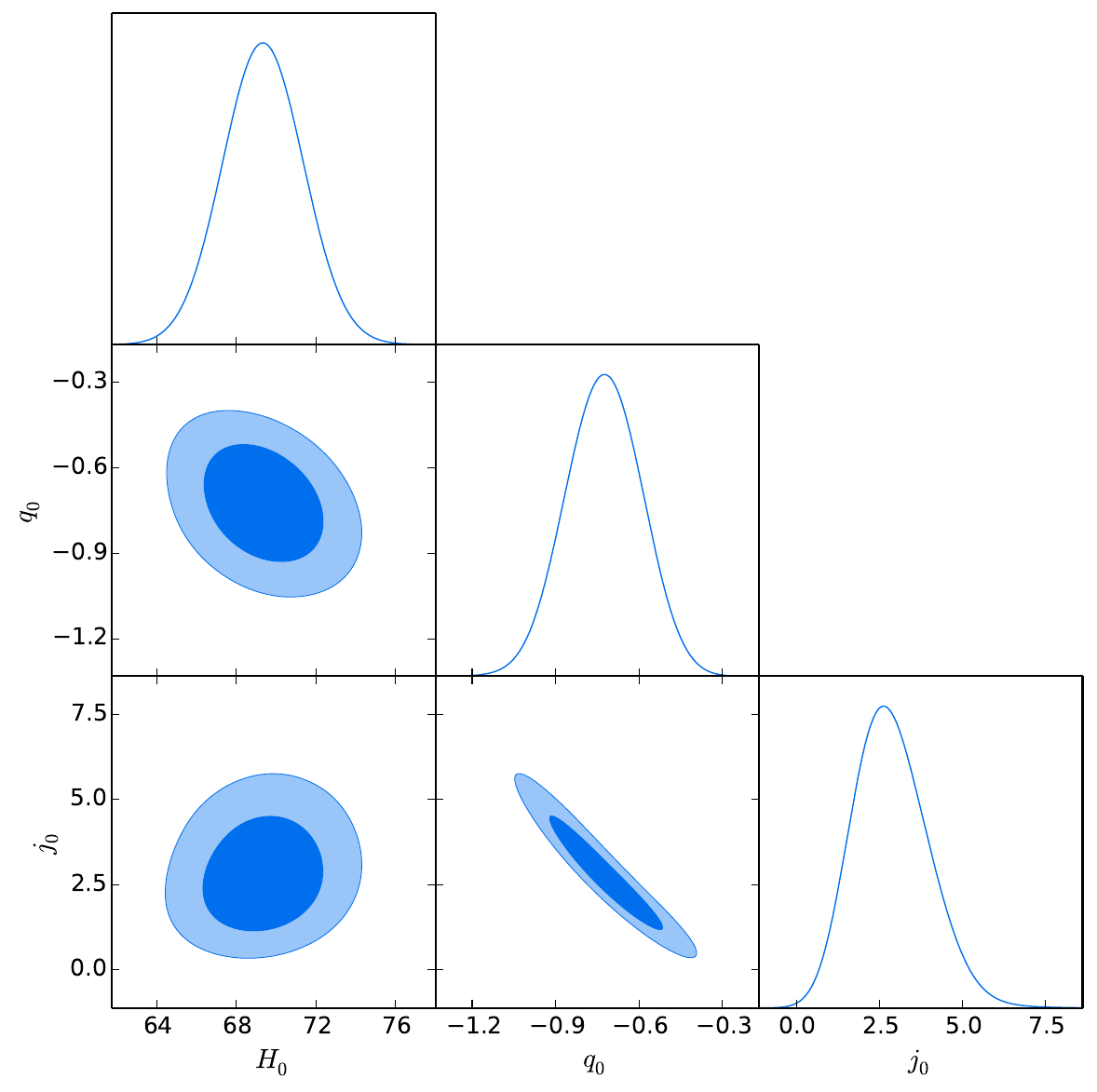}
    \caption{Marginalized contours at the 68\% and 95\% confidence levels, and posterior distributions for the cosmographic parameters as a result of our MCMC analysis.}
    \label{fig:triangle}
\end{figure}

For our purposes, we perform a Markov chain Monte Carlo (MCMC) analysis based on the Metropolis-Hasting algorithm \cite{Hastings70}. Assuming flat priors on the fitting parameters, our Bayesian analysis, applied to the combination of CC and SN data, provides us with the following results at the $1\sigma$ confidence level:
\begin{equation}
\left\{
\begin{aligned}
 H_0&=69.3^{+2.0}_{-2.0}\,, \\
 q_0&=-0.73^{+0.13}_{-0.13}\,, \\
 j_0&=2.84^{+1.00}_{-1.23}\,,
\end{aligned}
\right.
\label{constraints}
\end{equation}
where $H_0$ is expressed in units of km/s/Mpc. In \Cref{fig:triangle}, we show the $1\sigma$ and $2\sigma$ marginalized contours, with the posterior distributions, for the cosmographic parameters. 

\subsection{Evolution of the non-local fields}

Using \Cref{eq:dL_Pade21} with the mean values obtained in \eqref{constraints}, it is possible to frame the Universe evolution with high accuracy up to intermediate redshifts.
Thus, in our reconstruction procedure, we assume $8\pi G\rho=3H_0^2 h_{2,1}^2$, where $h_{2,1}\equiv H_{2,1}/H_0$ is the (2,1) Pad\'e  parametrization of the normalized Hubble rate. This can be expressed, for convenience, in terms of the  e-fold number, $N\equiv \ln a$:
\begin{equation}
h_{2,1}=\dfrac{\mathcal{P}(q_0,j_0)}{\mathcal{Q}(q_0,j_0)}\,,
\label{eq:H21}
\end{equation}
where 
\begin{subequations}
\begin{align}
&\mathcal{P}\equiv 4 e^{-4 N } \Big[e^{N } \left(j_0-3 q_0^2+2 q_0-2\right)-j_0+3q_0^2+q_0-1\Big]^4  \\
&\mathcal{Q}\equiv \Big\{14 + j_0 (7 - 10 q_0 - 9 q_0^2)+ 2 j_0^2 - 40 q_0 + 17 q_0^2 + 18 q_0^3\nonumber \\ 
&\hspace{0.8cm}+ 9 q_0^4 -2e^N \big( j_0  - q_0 -1 + 2 q_0^2+ 2 j_0^2 + 9 q_0^3 + 9 q_0^4 \nonumber \\ 
&\hspace{0.8cm}- 4 j_0 q_0  - 9 j_0 q_0^2\big)+e^{2N} \big[2 + 2 j_0^2 + 2 q_0 + 5 q_0^2 + 9 q_0^4	\nonumber \\
&\hspace{0.8cm} + j_0 (-5 + 2 q_0 - 9 q_0^2)\big]\Big\}^2.
\end{align}
\end{subequations}

On the other hand, one may describe the early cosmic times through $h_r^2=\Omega_{r0}e^{-4N}$, where $\Omega_{r0}\simeq 9.2 \times 10^{-5}$ is the current value of the radiation energy density estimated by the Planck collaboration \cite{Planck18}. Hence, assuming that matter behaves as dust, the only contribution to cosmic pressure is given by radiation, so that $8\pi G p=H_0^2 \Omega_{r0}e^{-4N}$.

The whole background history of the Universe could be thus parametrized as $h^2 \simeq h_r^2+h_{2,1}^2$, which allows us to take into account the  evolution from the early to the late epochs without the need to specify a particular cosmological model. 
While the early times are dominated by the radiation term, the latter becomes subdominant in recent times, implying $h\approx h_{2,1}$ for $|N|< 1$.

In order to study the cosmic evolution of  non-local fields, we first convert the time derivatives as
\begin{equation}
\frac{d}{dt}=H \frac{d}{dN}\,, \quad \frac{d^2}{dt^2}= H\left(H'\frac{d}{dN}+H \frac{d^2}{dN^2}\right),
\end{equation}
where the prime denotes the derivative with respect to $N$.
Then, \Cref{eq:master} becomes
\begin{equation}
W''+(5+\xi)W'+2(3+\xi)W=\mu\,,
\label{eq:W''}
\end{equation}
where $\xi\equiv h'/h$ and $\mu\equiv (3h_{2,1}^2-h_r^2)/h^2$.

Taking into account the expression for the Ricci scalar under the metric \eqref{eq:metric}, namely $R=-6(\dot{H}+2H^2)$, from \Cref{eq:X}, we find
\begin{equation}
X''+(3+\xi)X'+6(2+\xi)=0\,.
\label{eq:X''}
\end{equation}
Moreover, from \Cref{eq:Y}, one obtains 
\begin{equation}
Y''+(3+\xi) Y'={X'}^2\,,
\label{eq:Y''}
\end{equation}
whereas \Cref{eq:U} yields
\begin{equation}
U'+2 V X'=0\,.
\label{eq:U'}
\end{equation}
Finally, from \Cref{eq:V}, we find
\begin{equation}
V''+(3+\xi) V'+6 (2+\xi)\frac{df}{dY}=0\,,
\label{eq:V''}
\end{equation}
where $d f/d Y=(W'-U')/Y'$, with $Y'\neq 0$, as a consequence of the relation $f=W-U-1$.
We notice that the above differential equations are independent of the $H_0$ value.

As mentioned earlier, retarded boundary conditions are needed in order  to reproduce the late accelerated expansion of the Universe and avoid the presence of ghosts. 
For this reason, in performing  numerical calculations, we impose the following initial conditions in the radiation-dominated era, i.e., $N_0=-16$: $X_0=X_0'=U_0=V_0=V'_0=W_0=W'_0=0$.  As far as the scalar $Y$ is concerned, we choose $Y_0=0$, and $Y_0'=10^{-2}$ in order to avoid the divergence of the other fields due to \Cref{eq:V''}.

\begin{figure}
\includegraphics[width=3.2in]{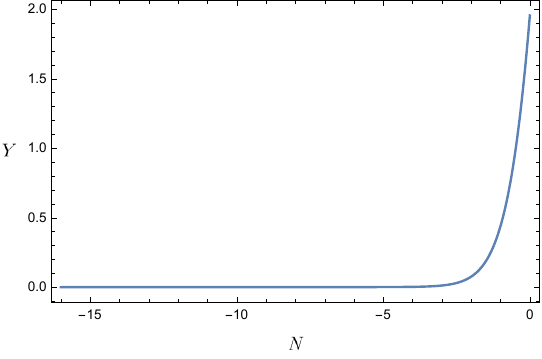}
\caption{Numerical reconstruction of the non-local field $Y$ as a function of the e-fold number.}
\label{fig:Y}
\end{figure}

\begin{figure}
\includegraphics[width=3.2in]{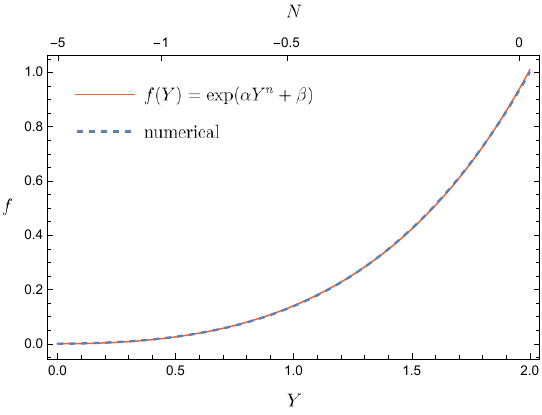}
\caption{Comparison between the numerical reconstruction of the non-local distortion function (dashed) and its best analytical approximation (solid), in terms of its natural argument, $Y$, and the e-fold number, $N$. The best-fit values of the free coefficients $\alpha$, $\beta$ and $n$ are given in \eqref{eq:coefficients}. }
\label{fig:f}
\end{figure}

\section{Results and discussion}
\label{sec:results}

The numerical solution to the system of \Cref{eq:W'',eq:X'',eq:Y'',eq:U',eq:V''} determines the dynamical behaviour of the auxiliary scalar fields. 

In \Cref{fig:Y}, we display the cosmic evolution of the non-local field $Y$ as inferred from our numerical reconstruction procedure. 
We notice that $Y$ is quiescent throughout the radiation-dominated epoch, while it grows during the matter dominance, becoming effective only at late cosmic times. A similar behaviour characterizes the other auxiliary scalars. 
Therefore, the effects of non-local fields are retarded to late times, thus providing a natural explanation for cosmic acceleration.

Moreover, after obtaining the solutions for $U$ and $W$, we determine the time evolution of the distortion function through $f=W-U-1$. 
Finally, combining the solutions for $f(N)$ and $Y(N)$, one can obtain the shape of $f(Y)$ defining the non-local action \eqref{action_nonlocal}.
In so doing, we find that the best analytical approximation to the numerical solution of $f(Y)$ is given by the function
\begin{equation}
f(Y)=\exp(\alpha Y^n+\beta)\,,
\label{eq:f_analytical}
\end{equation}
where the fitting coefficients take the following values:
\begin{equation}
\left(\alpha,\, \beta,\, n\right)=(11.5,\, 7.16,\, 0.23)\,.
\label{eq:coefficients}
\end{equation}

In \Cref{fig:f}, we show the numerical behaviour of $f(Y)$ compared to its best analytical approximation. One can clearly see that the outcome of our reconstruction procedure fully overlaps with the exponential function given by \Cref{eq:f_analytical}, throughout the entire cosmological evolution. In particular, the contribution of the distortion function to the gravitational action becomes significant during the most recent cosmic history, thus delaying the onset of accelerated expansion to late times.

\subsection{Comparison with previous findings}

We shall now discuss our results in light of those found in previous studies. Specifically, the reconstruction process described in \cite{Deser:2019lmm} showed that a simple form that fits the numerical solution for the distortion function is given by $f(Y)\approx e^{1.1(Y - 16.7)}$. 
It is worth noticing that this function matches the numerical results mainly at small $Y$, and it represents a simplified approximation of the full expression depending on the cosmological parameters of the $\Lambda$CDM scenario. 
In fact, although the resulting model was claimed to reproduce the current accelerated phase of the Universe expansion without resorting to the cosmological constant, however, the adopted numerical procedure is based on enforcing the $\Lambda$CDM expansion history. This fact might question the original purpose of overcoming the fine-tuning issue plaguing $\Lambda$.

On the other hand, our reconstruction method relies on a kinematic parametrization of the Hubble expansion rate, leading to bias-free results that are independent of \emph{a priori} assumptions on the cosmological background. Furthermore, the analytical form obtained in the present study reproduces with high accuracy the numerical evolution of $f(Y)$ over the whole cosmic domain.

\section{Summary and final remarks}
\label{sec:conclusions}

Inspired by the role of non-locality in extensions of GR to the ultraviolet scales, in the present study we considered a non-local cosmological scenario with the aim of explaining the accelerated expansion of the Universe without introducing dark energy. In particular, we investigated a modification of the Hilbert-Einstein action through the gravitational effects of the so-called distortion function.
After recasting the gravitational action in terms of auxiliary non-local scalar fields, we assumed the spatially flat FLRW geometry to analyze the cosmological behaviour of the model. Hence, we obtained the modified Friedman equations, where the density and the pressure of the cosmic fluid are deprived of the dark energy contribution. 

We presented a model-independent method to reconstruct the time evolution of the distortion function. In so doing, we parametrized the late cosmic history by means of kinematic variables arising from a Pad\'e approximation of the Hubble expansion rate. 
Then, including the radiation contribution to take into account the early epochs, we built up, from recent observations, a suitable parametrization of  cosmic evolution based on model-independent constraints. 

Therefore, adopting the mean result from our MCMC analysis, we solved the dynamical system describing the cosmological evolution of the auxiliary fields through retarded boundary conditions, which are required in order to reproduce the current cosmic acceleration and avoid ghost instabilities in the theory. In particular, we chose initial conditions such that the non-local scalars and their derivatives vanish in the early times, while showing their effects only at late times. 
We described the numerical reconstruction procedure to obtain the distortion function and, then, we inferred the analytical form that best fits the numerical solution over the whole cosmological domain. Our results show that the non-local fields are quiescent during the radiation era, while they start growing during the matter-dominated era. The effects of  distortion function become significant only recently, thus providing a natural explanation for the delay of the cosmic acceleration at late times.

In conclusion, we discussed our findings in view of previous predictions for the non-local model under study. Differently from earlier strategies, based on specific assumptions on the cosmological background, the method presented here is independent of any \emph{a priori} postulated expansion history or equation of state for the cosmic fluid. Thus, the present approach minimizes induced biases in the numerical procedure, and actually heals the fine-tuning issue of the cosmological constant that is at the origin of the proposed non-local gravity scenario.

\acknowledgments 
The authors acknowledge the support of Istituto Nazionale di Fisica Nucleare, Sezione di Napoli, {\it iniziativa specifica} QGSKY.
This paper is based upon work from COST Action CA21136 - Addressing observational tensions in cosmology with systematics and fundamental physics (CosmoVerse), supported by COST (European Cooperation in Science and Technology).

\appendix

\section{Datasets}
\label{sec:appendix}

In this appendix, we briefly describe the main features of  cosmic data we adopted   to constrain  cosmographic coefficients defining the (2,1) Pad\'e parametrization of the Hubble expansion rate.

Specifically, we combine the likelihood functions of cosmic chronometers (CC) and type Ia Supernovae (SN) data available from recent catalogs. Thus, to infer the mean values of our free parameters and their corresponding $1\sigma$ and $2\sigma$  confidence levels, we perform an MCMC analysis on the joint likelihood, $\mathcal{L}=\mathcal{L}_\text{CC}\times \mathcal{L}_\text{SN}$.

\subsection{Cosmic chronometers}

A reliable model-independent dataset, widely used to test cosmological scenarios, is based on the differential age method first proposed in \cite{Jimenez:2001gg}. 
Specifically, one can infer the value of the Hubble parameter by measuring the difference between the ages of pairs of nearby galaxies. The latter are passively evolving galaxies that act as cosmic chronometers, providing measurements of  cosmic expansion rate at $z\lesssim 2$ through the relation
\begin{equation}
    H(z)=-\frac{dz}{dt} (1+z)^{-1}\,.
\end{equation}
In our study, we make use of the uncorrelated measurements collected in \cite{Capozziello:2017buj} (and references therein), for which the likelihood reads
\begin{equation}
    \mathcal{L}_\text{CC} \propto \exp\left\{-\frac{1}{2}\sum_{i=1}^p\left[\frac{H_i^\text{(obs)}-H_i^\text{(th)}}{\sigma_{H,i}}\right]^2\right\} ,
\end{equation}
being $p$ the number of data points.

\subsection{Supernovae Ia}
The second dataset we employ in our analysis is the Pantheon catalog of SN Ia in the interval $0.01 < z < 2.3$ \cite{Pantheon}. Each SN is standardized by modeling the distance modulus as
\begin{equation}
    \mu=m_B - M +a x_1 -b c +\Delta_M +\Delta_B\,,
\end{equation}
where $M$ and $m_B$ are the SN absolute and apparent magnitudes, respectively, while $x_1$ is the stretch factor and $c$ is the color of the light curve. Moreover, $\Delta_M$ and $\Delta_B$ are corrections accounting for the host-mass galaxy and the distance bias, respectively, whereas $a$ and $b$ are free coefficients.

The above parametrization can be confronted with the theoretical distance modulus defined in terms of the luminosity distance relative to the cosmological model under consideration:
\begin{equation}
    \mu(z)=25+5\log\left[\frac{d_L(z)}{1\ \text{Mpc}}\right].
\end{equation}

As shown in \cite{Riess:2017lxs}, one may use the Pantheon datasets to constrain the normalized Hubble parameter in a model-independent way by means of the likelihood function
\begin{equation}
    \mathcal{L}_\text{SN}\propto\exp\left(-\frac{1}{2}\textbf{V}^\text{T} \Sigma^{-1}\textbf{V}\right),
\end{equation}
where $\textbf{V}=(v_1,\hdots,v_p)$, with $v_i=h_{i,\text{obs}}^{-1}-h^{-1}_{i,\text{th}}$, for $i=1,\hdots, p$, being $p$ the number of data points. Here, $\Sigma$ is the covariance matrix as given in \cite{Riess:2017lxs}.

{}

\end{document}